\documentclass[letterpaper,conference]{IEEEtran}
\IEEEoverridecommandlockouts

\usepackage{tikz}

\newcommand\copyrighttext{%
  \footnotesize \textcopyright 2021 IEEE. Personal use of this material is permitted. Permission from IEEE must be obtained for all other uses, in any current or future media, including reprinting/republishing this material for advertising or promotional purposes, creating new collective works, for resale or redistribution to servers or lists, or reuse of any copyrighted component of this work in other works.}
\newcommand\copyrightnotice{%
\begin{tikzpicture}[remember picture,overlay]
\node[anchor=south,yshift=10pt] at (current page.south) {\fbox{\parbox{\dimexpr\textwidth-\fboxsep-\fboxrule\relax}{\copyrighttext}}};
\end{tikzpicture}%
}

\newcommand\acceptedtext{%
  \footnotesize Accepted at The 2021 IEEE 46th Conference on Local Computer Networks (LCN)}
\newcommand\acceptednotice{%
\begin{tikzpicture}[remember picture,overlay]
\node[anchor=north,yshift=-30pt] at (current page.north) {\fbox{\parbox{\dimexpr\textwidth-\fboxsep-\fboxrule\relax}{\acceptedtext}}};
\end{tikzpicture}%
}

\usepackage{cite}
\usepackage{amsmath,amssymb,amsfonts}
\usepackage{algorithmic}
\usepackage{graphicx}
\usepackage{textcomp}
\usepackage{xcolor}
\def\BibTeX{{\rm B\kern-.05em{\sc i\kern-.025em b}\kern-.08em
    T\kern-.1667em\lower.7ex\hbox{E}\kern-.125emX}}

\graphicspath{{figures/}}

\ifCLASSOPTIONcompsoc
	\usepackage[caption=false,font=normalsize,labelfont=sf,textfont=sf]{subfig}
\else
	\usepackage[caption=false,font=footnotesize]{subfig}
\fi

\begin{document}

\title{On the Analysis of Adaptive-Rate Applications in Data-Centric Wireless Ad-Hoc Networks
}

\author{\IEEEauthorblockN{Md Ashiqur Rahman}
\IEEEauthorblockA{
	\textit{The University of Arizona}\\
	marahman@cs.arizona.edu}
\and
\IEEEauthorblockN{Beichuan Zhang}
\IEEEauthorblockA{
	\textit{The University of Arizona}\\
	bzhang@cs.arizona.edu}
}

\maketitle

\copyrightnotice
\acceptednotice

\vspace{-10pt}
\begin{abstract}
 Adapting applications' data rates in multi-hop wireless ad-hoc networks is inherently challenging. Packet collision, channel contention, and queue buildup contribute to packet loss but are difficult to manage in conventional TCP/IP architecture. This work explores a data-centric approach based on Name Data Networking (NDN) architecture, which is considered more suitable for wireless ad-hoc networks. We show that the default NDN transport offers better performance in linear topologies but struggles in more extensive networks due to high collision and contention caused by excessive Interests from out-of-order data retrieval and redundant data transmission from improper Interest lifetime setting as well as in-network caching. To fix these, we use round-trip hop count to limit Interest rate and \textit{Dynamic Interest Lifetime} to minimize the negative effect of improper Interest lifetime. Finally, we analyze the effect of in-network caching on transport performance and which scenarios may benefit or suffer from it. 
\end{abstract}


\section{Introduction}
\label{sec:introduction}

Communication in a wireless environment is complex by nature because of channel access, contention, packet collision, signal interference, to name a few \cite{goldsmith2005wireless}. A multi-hop network makes it more challenging. A mobile ad-hoc network (MANET) exacerbates the packet loss even further as the uncertainty of link breakage between two neighbor nodes increases manifold. Thus, the application-level performance is much lower than in a stable wired network.

A transport-layer protocol ensures end-to-end data delivery and adapts data rate to the available bandwidth in the network, such that applications can fully utilize available bandwidth without causing network congestion. Moreover, sending rate or congestion window also affects wireless contention and throughput \cite{fu2003impact}. Today's IP-based transport protocols for adaptive-rate applications based on TCP and its variants assume point-to-point data transportation over a relatively stable path. Mobility, however, often breaks this assumption as end-to-end paths break more often. It results in significant RTT variance, out-of-order data arrival, and higher packet loss. Thus, rate management becomes tricky for transport. 

A data-centric communication moves away from the point-to-point abstraction of IP. Here, the network layer can deliver or retrieve data from anywhere in the network, even from multiple data points at the same time. As a result, a data-centric transport can support out-of-order data arrival. On the other hand, the in-order delivery of TCP and its variants have many limitations in MANETs~\cite{holland2002analysis}, e.g., head-of-line blocking, which yields low throughput. In theory, a data-centric transport should perform better than traditional TCP/IP in an ad-hoc wireless environment. 

One such data-centric architecture is the emerging Named Data Networking or NDN~\cite{ndn-ccr}. In NDN, requests (called Interest packets) and replies (called Data packets) carry data \textit{name} instead of IP addresses. The function of the network is to retrieve named data instead of delivering packets to a particular node. Thus data can be supplied from any node, arrive in any order, and via any path. As a result, NDN natively supports out-of-order data retrieval, multicast, multihoming, and in-network caching. It also has a \textit{stateful forwarding plane}, enabling fast detection of network anomalies and recovering from them accordingly. However, due to the inherent challenges of the wireless environment, it is unclear how much NDN's architectural advantage would translate into actual performance gain and how much optimization or additional engineering would be needed to maximize it even further.

This work analyzes the transport and forwarder behavior in data-centric adaptive-rate applications using NDN in wireless ad-hoc networks. We start with a proof-of-concept using a simple AIMD-based window adaptation. It shows that NDN can outperform TCP in both wired and static wireless linear topology under lossy conditions with a single sender-receiver (IP) or consumer-producer (NDN) pair. However, in a more extensive scenario with multiple consumer-producer pairs, NDN's out-of-order delivery causes abrupt congestion-window adjustment, degrading its throughput. We identify channel contention as a root cause with almost no congestion at the device queue. We verify and overcome such behavior by applying a congestion window limit (CWL)~\cite{chen2003settingCWL}. NDN offers better throughput with this change than TCP/IP under mobility with caching and multicast utilization.

Next, an NDN consumer application adds a randomly generated \verb|NONCE| with each Interest irrespective of original or retransmission and helps to detect in-network loops. The consumer application also adds a lifetime to the Interests (usually fixed, e.g., 2s). It helps achieve better in-network multicast by aggregating Interests with the same name and different \verb|NONCE|s in the Pending Interest Table (PIT). However, using a fixed lifetime in an adaptive-rate application can cause PIT aggregation on consumer retransmissions (different \verb|NONCE|s). Consequently, NDN multicast leads to redundant data over multiple paths to a consumer, causing high collision and contention. We propose a novel Dynamic Interest Lifetime (DIL) mechanism based on the application's round-trip timeout (RTO) and use a multiplier to delay the timeout event check to negate this effect. It allows the network to clear out stale PIT entries on application retransmission while keeping the aggregation opportunity if multiple consumers ask for the same data within a close interval.

Finally, NDN caching offers better resiliency to packet loss. However, it can also increase redundant data transmissions in the network from disjoint paths. Such transmission overhead inherently leads to increased packet collision and channel contention. Thus we show that caching in some ad-hoc communication scenarios might not be as beneficial.

Through this work, we confirm that a data-centric transport like in NDN improves adaptive-rate application's performance, especially in small lossy networks. In an extensive wireless network, however, the data-centric mechanisms such as out-of-order delivery, multicast, in-network caching can have undesirable side effects that would reduce throughput. We then present meaningful insights into these critical challenges of developing data-centric adaptive-rate applications in a wireless ad-hoc network and propose solutions to mitigate them for achieving better throughput. Our results call for more research into new mechanisms and designs that can take advantage of the data-centric architecture in a wireless environment to achieve a breakthrough in overall performance.

\section{Related Works}
\label{sec:related-works}

The IP transport layer supports minimal loss recovery as it is decoupled from the network layer. Thus, \cite{dyer2001comparison, chlamtac2003mobile} show how TCP struggles to perform well in mobile wireless networks from well-known head-of-line blocking and packet loss from pushing data in the network. The idea is to maintain a \textit{full pipe} with in-flight packets even \textit{before} receiving acknowledgments for sent data to achieve a throughput close to the theoretical maximum. However, in a wireless network, with or without mobility, a push-based model causes a high amount of packet loss and retransmissions.

The TCP header limits loss recovery even with \texttt{SACK}~\cite{mathis1996tcp} and adjusts window size on both \texttt{ACK}/\texttt{SACK}, timeouts and explicit congestion notification. Thus adapting TCP in ad hoc networks requires substantial modifications such as \cite{sundaresan2005atp} or enabling out-of-order data retrieval \cite{wang2002improving}. More recent protocols like QUIC~\cite{langley2017quic} achieve per-packet loss detection and out-of-order retrieval in web applications showing the networking trend moving towards data-centric communication.

\cite{rahman2021datacentric} shows that NDN offers a better network forwarder than IP in MANETs but only considers constant bit-rate (CBR) traffic, excluding transport analysis. Furthermore, NDN's pull-based model controls the consumer's Interest sending rate as a Data node only replies on Interest packets. As an Interest packet is much smaller than a Data packet, one can assume that Interest loss from a high sending rate is far less precarious. 
However, in a multi-hop wireless scenario, it can still contribute to channel contention. \cite{amadeoNdnTransport2014} proposes rate-based transport for NDN using the inter-data gap (IDG). However, mobility may lead to frequent data node switching, such that a consumer may not get enough data packets for the sampling process. Their data node id-based design also deviates from a complete decentralized data-centric forwarding. \cite{bouk2015dpel} proposes a dynamic PIT entry lifetime (DPEL) based on Interest satisfaction rate and hop count. However, the hop count between a consumer and data node will likely fluctuate frequently in a dynamic network. As a result, PIT entries can timeout while data comes back when a data node is further away.

\section{Motivation}
\label{sec:motivation}
\subsection{Out-of-order Retrieval Offers Better Transport}
\label{sec:out-of-order-retrieval}

\begin{figure}[!t]
  \centering
  \subfloat[TCP/IP]{
    \includegraphics[width=0.45\columnwidth]{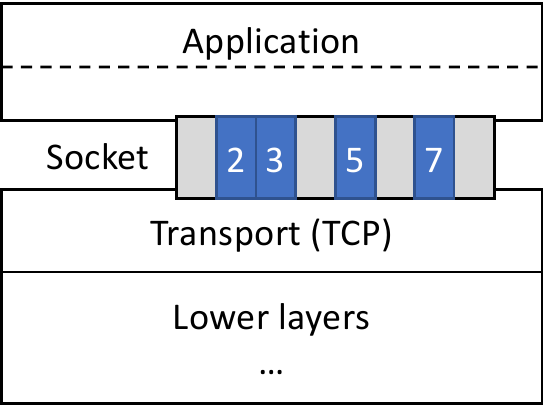}
    \label{fig:tcp-transport}
  }
  \hfill
  \subfloat[NDN]{
    \includegraphics[width=0.45\columnwidth]{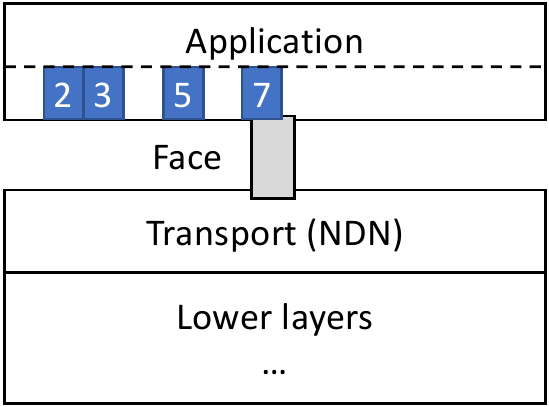}
    \label{fig:ndn-transport}
  }
  \caption{Packet flow from lower layers to the application in TCP/IP and NDN.}
  \label{fig:transport-comparison}
\end{figure}

A fundamental difference between data-centric networking and traditional TCP/IP is how data packets are forwarded from the lower layers to the uppermost application layer. Data-centric communication supports out-of-retrieval by default, while IP requires specialized protocols such as \cite{wang2002improving} for the same purpose as TCP assumes in-order delivery. Thus to understand the transport benefits of data-centric networking, we show a side-by-side comparison between the NDN transport (or consumer node's forwarder) and traditional TCP/IP byte-stream management in Fig.~\ref{fig:transport-comparison}.

In Fig.~\ref{fig:tcp-transport}, the socket provides an interface between the transport and application layer. It also offers a packet buffer to manage the in-order delivery. The example also shows the head-of-line blocking for packets 1 to 7 because packet 1 was either lost in the network or traveling through a possible longer path (multipath or path change). As a result, even though packets 2 and 3 arrived earlier, they will be buffered until the first arrives. Selective acknowledgment or \texttt{SACK} \cite{mathis1996tcp} can help TCP with better loss recovery but is limited to maximum three-hole detection in the byte stream. Such limitation comes from only using the fixed-sized TCP header and not a variable-length design for performance reasons.

NDN's \textit{\textbf{face}} in Fig.~\ref{fig:ndn-transport} is similar to the socket interface and sits between the transport and the application. However, it does not buffer any data packet from the transport; instead, it relays to the application right away, responsible for packet ordering. Such behavior avoids possible buffer overflow like in TCP socket and provides infinite \texttt{SACK} with per-Interest-Data communication. 

Moreover, out-of-order delivery in TCP would mean packet loss in the network. In wired networks, it is highly likely from a congestion drop. In wireless networks, however, it can also be from packet collision, channel contention or path breakage from mobility. In data-centric networking, the out-of-data retrieval indicates that the network communication is available, irrespective of multipath, congestion, or channel contention. Later, a retransmitted Interest can also retrieve cached data. Therefore out-of-order retrieval enables NDN to advance the congestion window on each data packet, while TCP/IP does not do so by default. Thus data-centric transport promises better performance. 
\subsection{Proof-of-concept for Data-centric Transport}
\label{sec:ndn-default-transport}

We now show a proof-of-concept on the potentials of using data-centric communication at the transport level without any complex engineering as in IP's host-centric model. It also shows the advantages of NDN's out-of-order data retrieval under lossy conditions. To do so, we consider the well-known TCP-NewReno \cite{tcpNewRenoRFC} for TCP/IP. An AIMD algorithm for NDN applications follows the same slow-start and congestion avoidance for congestion window ($cwnd$) maintenance. 
It starts with an initial congestion window or $cwnd=1$ and an initial threshold, $ssthresh=\infty$. On each data packet, the slow start or congestion avoidance for $cwnd$ works as,
\begin{equation}
  cwnd += 
\begin{cases}
  1.0 & \text{if } cwnd < ssthresh \text{ (slow start)}\\
  \frac{1.0}{cwnd}, & \text{otherwise (congestion avoidance).}
\end{cases}
\label{eq:ndn-aimd-increase}
\end{equation}

The NDN AIMD also uses the conservative window adaptation (CWA) \cite{blantonTcpSack} on consumer application's timeout followed by the well-known TCP retransmission timeout (RTO) calculation \cite{TCP-RTx}. The difference being for CWA, the Interest packet replaces TCP byte-stream, and the Data packet replaces TCP \texttt{ACK}. Thus, on an application timeout, if the CWA conditions meet, the multiplicative decrease in NDN is as follows,
\begin{equation}
  ssthresh = cwnd * \beta
\label{eq:ndn-aimd-decrease-thresh}
\end{equation}
\vspace{-5mm}
\begin{equation}
  cwnd = max(ssthresh, 1).
\label{eq:ndn-aimd-decrease-cwnd}
\end{equation}
Here, similar to the TCP-NewReno, $\beta=0.5$ in Eq.\ref{eq:ndn-aimd-decrease-thresh}, while Eq.\ref{eq:ndn-aimd-decrease-cwnd} ensures $cwnd$ never falls below the initial window size. For explicit congestion control, IP uses explicit congestion notification (ECN) while NDN uses congestion marking (CM) directly with returning data packets for outgoing queue build-up as described in \cite{KlausPcon}. 

With this NDN AIMD design, we divide the proof-of-concept for transport behavior analysis into two scenarios, 1) linear wired topology and 2) static, linear wireless topology.

\begin{figure}[!ht]
  \centering
  \subfloat[Throughput without bottleneck.]{
    \includegraphics[width=0.47\columnwidth]{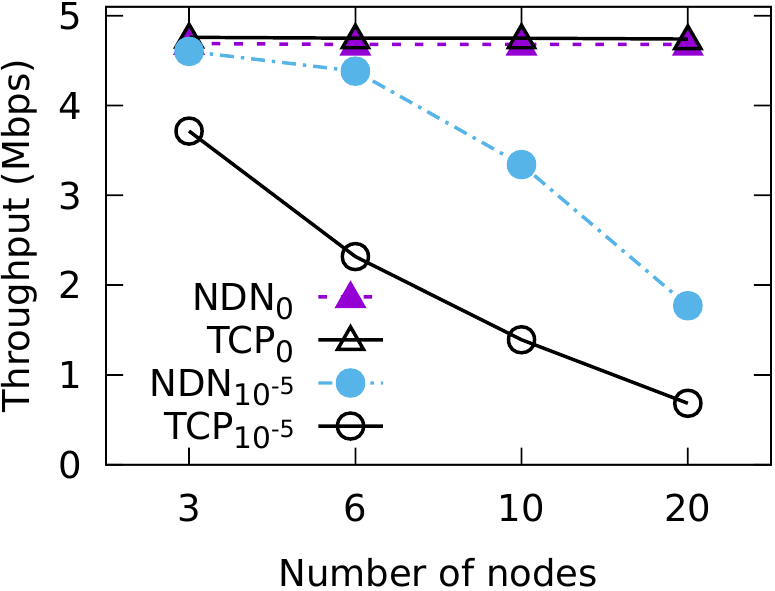}
    \label{fig:th-err-nobtl}
  }
  \hfill
  \subfloat[Throughput with bottleneck.]{
    \includegraphics[width=0.47\columnwidth]{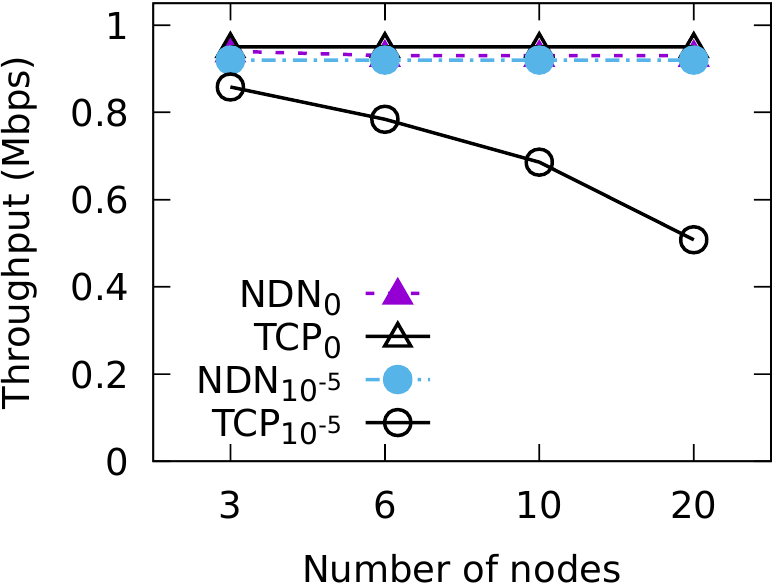}
    \label{fig:th-err-btl}
  }
  \par
  \subfloat[CWND without bottleneck.]{
    \includegraphics[width=0.47\columnwidth]{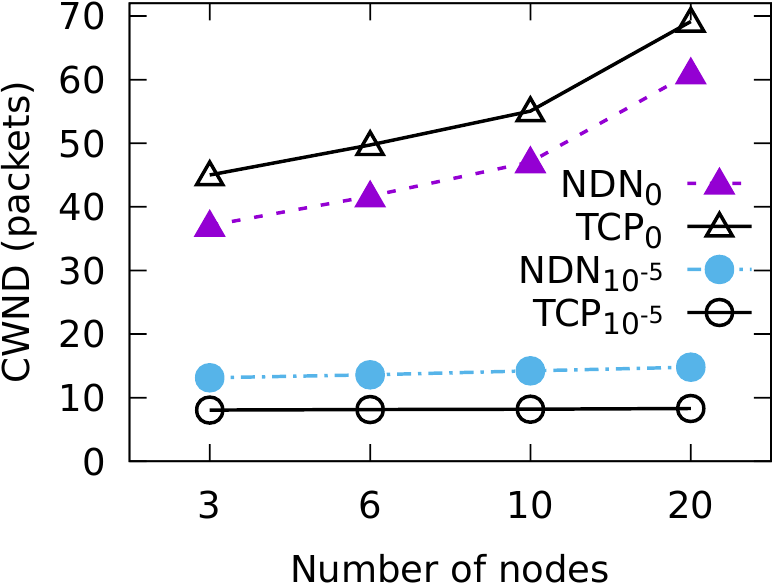}
    \label{fig:cwnd-err-nobtl}
  }
  \hfill
  \subfloat[CWND with bottleneck.]{
    \includegraphics[width=0.47\columnwidth]{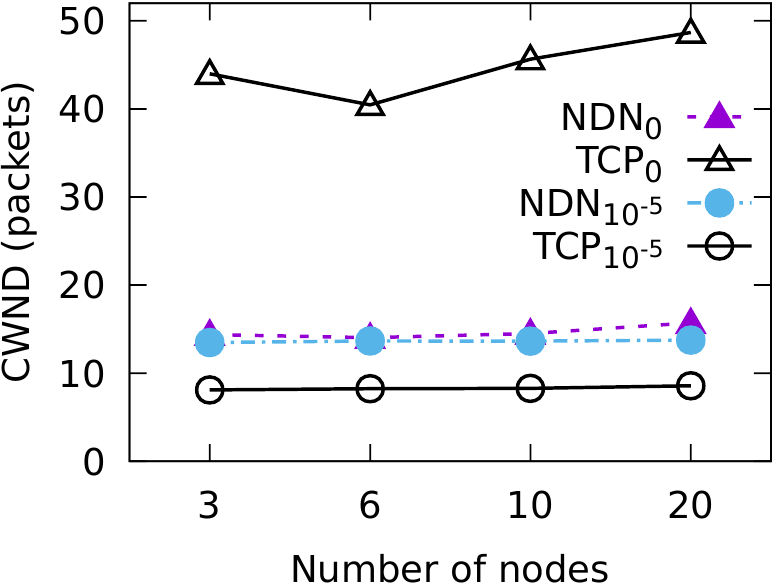}
    \label{fig:cwnd-err-btl}
  }
  \par
  \subfloat[RTT without bottleneck.]{
    \includegraphics[width=0.47\columnwidth]{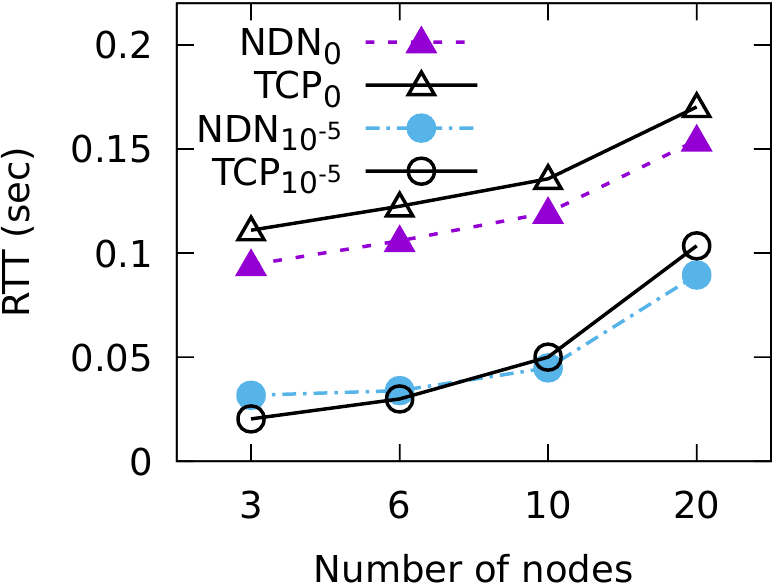}
    \label{fig:rtt-err-nobtl}
  }
  \hfill
  \subfloat[RTT with bottleneck.]{
    \includegraphics[width=0.47\columnwidth]{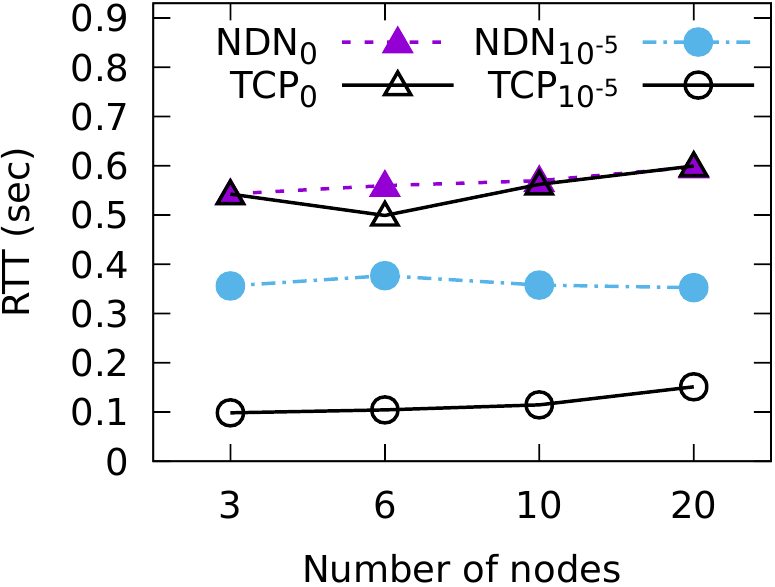}
    \label{fig:rtt-err-btl}
  }
  \caption{Effect of packet error and number of nodes on throughput, congestion window (CWND), and round-trip time (RTT) with and without bottleneck link in a linear wired topology. Subscript shows per-byte corruption probability = 0 or $10^{-5}$ ($\approx$1.5\% per-packet) at the sender and receiver interfaces.}
  \label{fig:wired-transport}
\end{figure}

\subsubsection{Linear wired topology}
\label{subsubsec:wired-proof-of-concept}
We simulate single consumer-producer (NDN) or server-client (TCP/IP) communication over different wired chain lengths using ndnSIM~\cite{mastorakis2017evolution}. We also consider per-byte error (or loss) probability at the end nodes' interfaces to emulate loss from wireless collision and contention. Furthermore, we test with and without bottleneck links in the chain to observe congestion signals. By default, all links are 5 Mbps with a 1 ms propagation delay. We set only one link in the middle as 1 Mbps with a 10ms propagation delay when the bottleneck is enabled. Each data packet payload is 1460 bytes. We collect an average of five 300 second runs and show the results in Fig.~\ref{fig:wired-transport}.

\textbf{Without bottleneck:} TCP and NDN have similar throughput without packet loss in Fig.~\ref{fig:th-err-nobtl}. However, under lossy conditions, their throughput go down with increasing chain length, with TCP performing the worse. TCP tries to keep the socket buffer full and yields high CWND (Fig.~\ref{fig:cwnd-err-nobtl}) and RTT (\ref{fig:rtt-err-nobtl}) when there is no loss. With packet loss, the in-order byte-stream cannot increase CWND due to SACK limitations (max. three-hole filling) in the header and socket buffer capacity. The resulting low RTT verifies our claims. On the other hand, NDN mainly lowers CWND on application timeout as there is minimal congestion marking without a bottleneck. Under packet loss, its infinite-SACK keeps the CWND higher than TCP, helping to achieve higher throughput.

\textbf{With bottleneck:} TCP and NDN show similar throughput in Fig.~\ref{fig:th-err-btl} without error. However, with error, only TCP's throughput degrades over chain length. Despite ECN, the SACK limitation keeps the CWND very low (Fig.~\ref{fig:cwnd-err-btl}), which we verify with the low RTT in Fig.~\ref{fig:rtt-err-btl}. NDN maintains a stable CWND even with packet error using CM and out-of-order data. Thus throughput change is minimal over chain length.

\begin{figure}[!t]
  \vspace{1mm}
  \centering
  \includegraphics[width=0.85\columnwidth]{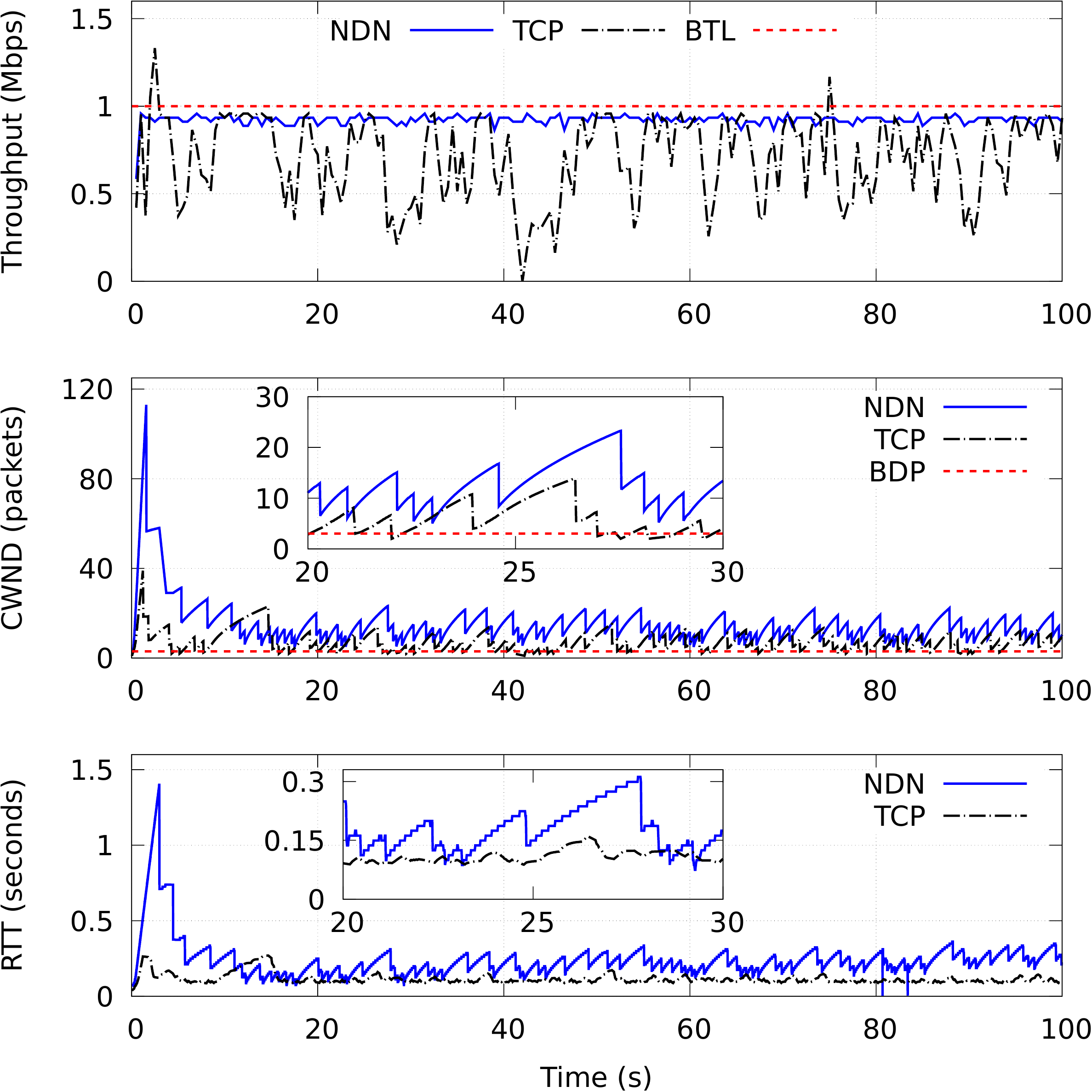}
  \caption{Transport behavior over time in a single simulation with per-byte loss probability = $10^{-5}$ ($\approx$1.5\% per-packet) at the sender and receiver interfaces with bottleneck-link = 1 Mbps in 10 nodes linear wired topology. Red lines in Throughput and CWND show the ideal bottleneck (BTL) and bandwidth-delay product (BDP), respectively.}
  \label{fig:ndn-tcp-err-effect-btl}
\end{figure}

Fig.~\ref{fig:ndn-tcp-err-effect-btl} shows the application throughput, $cwnd$, and RTT over the first 100 seconds in a single simulation and further verifies the advantages of NDN and shortcomings of TCP under lossy conditions and bottleneck. We see that TCP shows high fluctuations in throughput because of in-order delivery. As a result, application throughput goes above the link capacity at time = 2s and 75s as it processes more bytes from the receiver socket than the bottleneck capacity. On the other hand, NDN maintains a stable throughput over time with out-of-order retrieval and maintains a higher CWND and RTT, as discussed earlier. NDN's CWND and RTT spikes at the beginning verify the out-of-order effect with an initial RTO of 2 seconds (TCP reacts comparatively early with \texttt{ACK}/\texttt{SACK}).

\begin{figure}[!t]
  \centering
  \subfloat[Throughput.]{
    \includegraphics[width=0.47\columnwidth]{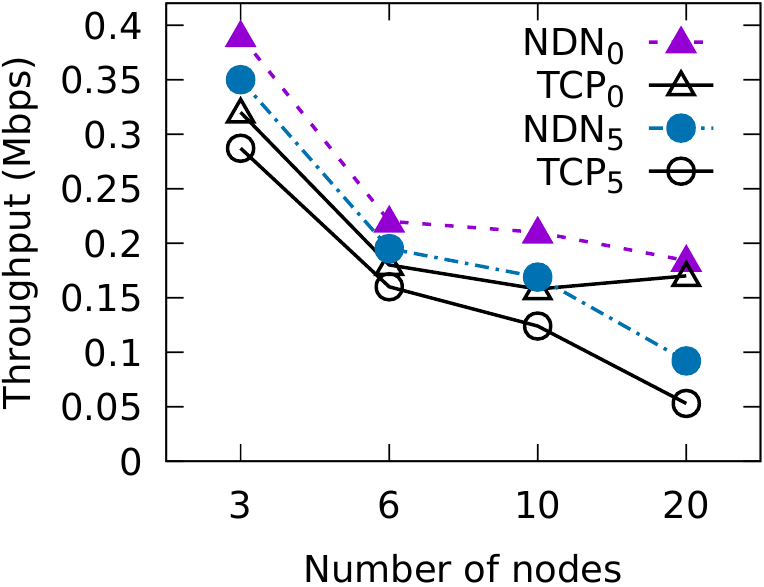}
    \label{fig:wl-thr-err}
  }
  \hfill
  \subfloat[CWND and RTT.]{
    \includegraphics[width=0.47\columnwidth]{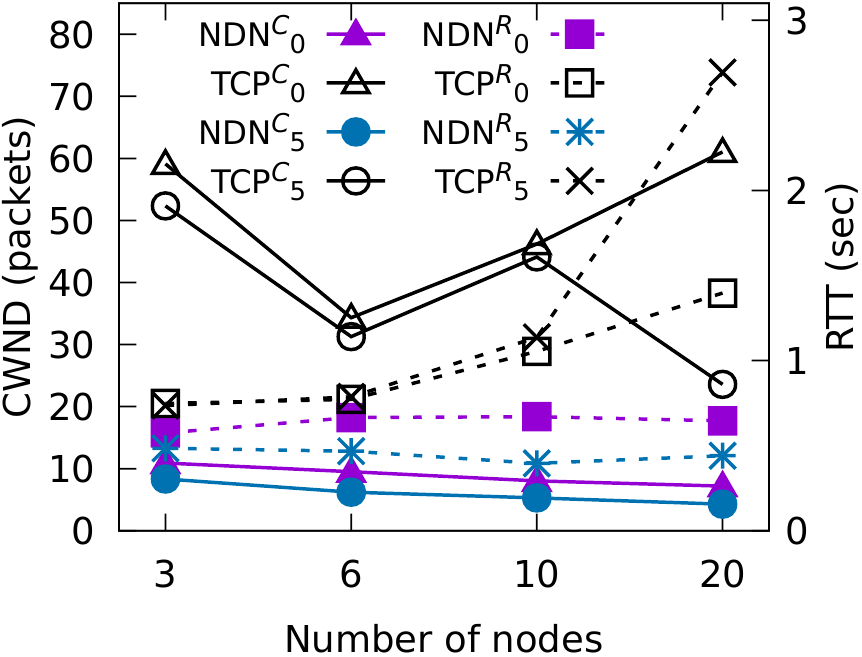}
    \label{fig:wl-cwnd-rtt-err}
  }
  \caption{Effect of packet error and number of nodes on throughput, congestion window (CWND), and round-trip time (RTT) in a static linear wireless topology. Superscript shows CWND=$C$ or RTT=$R$. Subscript shows per-packet loss probability percentage at each node's wireless interface.}
  \label{fig:wireless-transport}
\end{figure}

\subsubsection{Static, linear wireless topology}
\label{subsec:wireless-linear}

We use a static linear topology similar to Sec.~\ref{subsubsec:wired-proof-of-concept} for different chain lengths in multi-hop static wireless networks. However, each node has a single wireless interface covering a maximum of two neighbors. They communicate using IEEE 802.11b over a single channel at 1 Mbps with RTS/CTS disabled and 512 bytes payload. We collect five runs of 300 seconds per chain length. Each node's interface has a queue of 25 packets with a packet reception error rate of 0\% or 5\% to emulate added loss (e.g., signal jamming in military networks). Fig.~\ref{fig:wireless-transport} shows the NDN and IP transport behavior.

We can see that both TCP and NDN's maximum throughput are lower than their wired scenario and sharply drops with increasing chain length. It is because channel contention starts taking over NIC queue congestion, and the hidden terminal problem is also possible from bi-directional traffic. Our simulation logs also show only a few congestion signals. As a result, NDN controls $cwnd$ primarily based on application timeout, while TCP does so with \texttt{ACK}/\texttt{SACK} and timeout. 

TCP still tries to keep the receiver socket buffer full, which leads to high $cwnd$ and RTT at the sender in Fig.~\ref{fig:wl-cwnd-rtt-err} with and without additional packet error. However, TCP's in-order delivery leads to lower throughput than NDN's out-of-order retrieval in Fig.~\ref{fig:wl-thr-err}. A single simulation test also showed CWND and RTT spikes at the beginning (separate plot not shown), similar to Fig.~\ref{fig:ndn-tcp-err-effect-btl}, verifying NDN's out-of-order retrieval. However, they both go below TCP's as there are no separate buffers in NDN transport. As a result, NDN reacts to timeouts much faster compared to TCP. NDN shows an average of 21.31\% and 38.42\% more throughput than TCP/IP with 0\% and 5\% packet error, respectively.

\subsection{Data-centric Forwarding Improves Network Performance}
\label{sec:network-forwarding}

The data-centric ad-hoc forwarding (DAF)~\cite{rahman2021datacentric} shows that NDN can be a better fit than IP in a MANET environment while keeping the similar characteristics of broadcast-based learning, like the AODV~\cite{Perkins:1999:AODV} routing. DAF significantly reduces the retrieval latency, network load and thus improves the retrieval success rate. It does so by allowing any node to fall back to discovery for quick mobility and loss reaction. Each node maintains a weighted moving average of the RTT for each producer prefix and next-hop pair. Thus any node can expire its Forwarding Information Base (FIB) entry using the same TCP-RTO principle \cite{TCP-RTx}.

\begin{figure}[!t]
  \vspace{1mm}
  \centering
  \includegraphics[width=0.98\columnwidth]{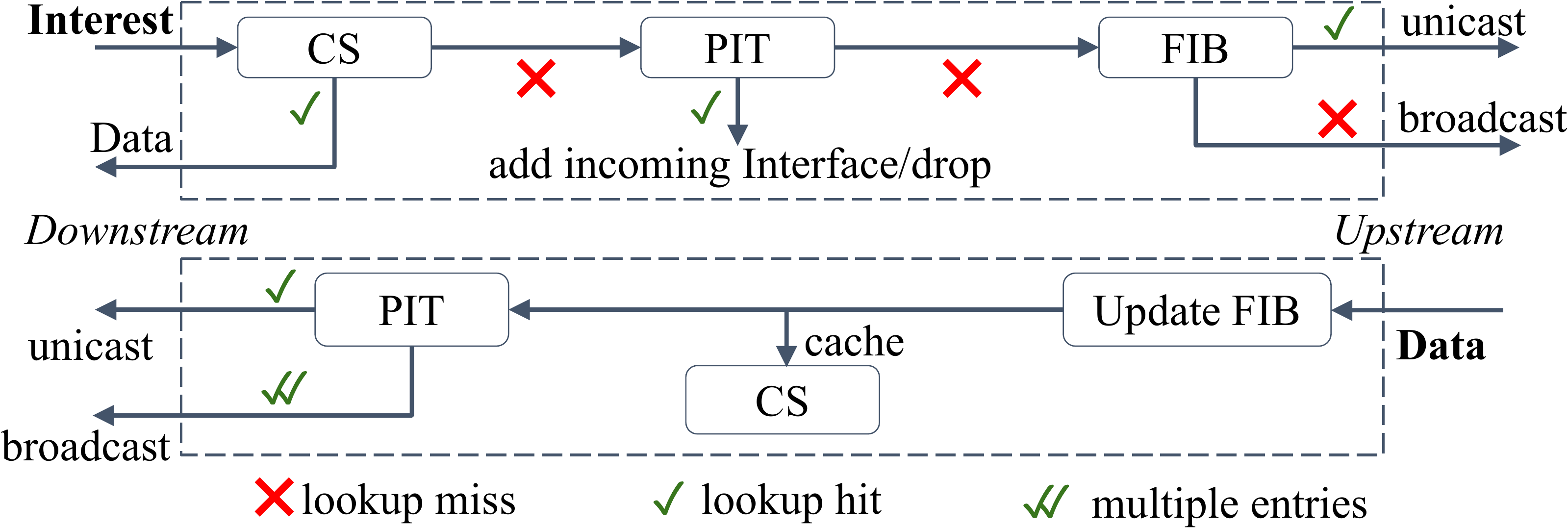}
  \caption{Forwarding pipeline of the DAF strategy for MANETs.}
  \label{fig:daf-forwarding}
\end{figure}

The forwarding pipeline of DAF is in Fig.~\ref{fig:daf-forwarding}. A node receiving an Interest first checks the Content Store (CS) and only returns data on a cache hit; otherwise, it forwards to the Pending Interest Table (PIT). The PIT aggregates and suppresses an Interest if it finds the same name with a different \texttt{NONCE} or drops if the same \texttt{NONCE} (loop); otherwise, forwards the unique Interest to the FIB. A node then unicasts the Interest to a valid FIB entry and broadcasts otherwise. 

A node receiving a Data packet on the downstream first updates its FIB with the sender information and then opportunistically caches the data to the CS. It benefits the network under mobility with uncertain link breakage. A node then performs a PIT lookup and unicasts data if there is one downstream, broadcasts on two or more entries to reduce transmission overhead, and drops if no entry is found. 

Moreover, DAF avoids negative acknowledgments or \texttt{NACK}s to reduce network traffic and allows a node to broadcast data for aggregated Interests in the PIT to avoid multiple data transmissions. Later in this paper, we use the DAF and AODV for NDN and IP, respectively, in wireless ad-hoc simulations. It ensures that the network level forwarding behavior is similar to provide a fair comparison at the transport level. 

\section{System model}
\label{sec:system-model}

\begin{figure}[!t]
  \vspace{1mm}
  \centering
  \includegraphics[width=0.90\columnwidth]{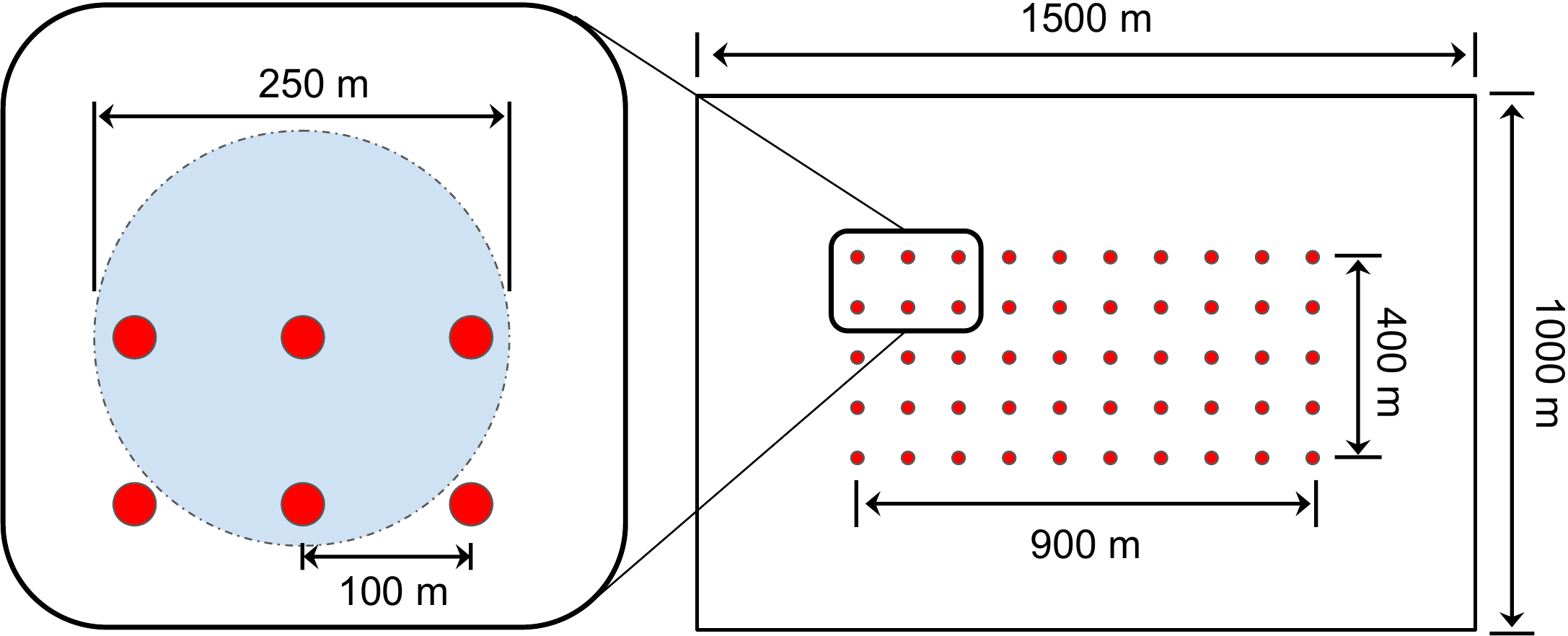}
  \caption{Initial system setup with stationary nodes. Right side shows the grid topology when nodes are stationary, and left side shows inter-node distance and transmission range (dotted large circle).}
  \label{fig:system-example}
\end{figure}

We use the random geometric graph~\cite{dall2002random} to analyze NDN's adaptive-rate applications in wireless networks. We randomly place 50 nodes in a 10x5 grid setup with 100 meters of separation in X and Y coordinates and 250 meters (diameters) transmission range for each node. Therefore, one node can communicate to a maximum of four nodes at any given time. It allows us to analyze the transport and network behavior when nodes are stationary, with a valid path between a consumer and producer. The simulation area is 1500m x 1000m, and the system model topology is in Fig.~\ref{fig:system-example}.

With mobility (speed $>0$ m/s), nodes start to move around at time $t>0$ using a random-walk 2D mobility model with 5 seconds duration for each randomly selected direction and no pause before selecting the next one. Speed ranges between 0 and 8 m/s, and in each simulation, all nodes maintain a constant speed. Each node has similar NIC properties as in Sec.~\ref{subsec:wireless-linear}. The network includes ten consumers or clients, 10 or 2 producers or servers, while the rest act as packet forwarders. The purpose of such mobility is to solely focus on the transport behavior in a generalized MANET scenario, and unlike in more specialized cases such as vehicular networks where mobility direction may play a vital role. We expect future research to tackle such design considerations.

\section{Data-centric Adaptive-rate Applications}
\label{sec:adtaptive-rate-ndn-manet}

With NDN's advantages at the application, transport, and network layer, we now analyze the behavior of data-centric adaptive-rate applications in a more extensive wireless network with multiple consumers, producers, and mobility. 

\subsection{Channel Contention from Out-of-order Data Retrieval}
\label{subsec:channel-contention}

In Sec.~\ref{subsec:wireless-linear}, we saw that wireless communication induces channel contention, overshadowing queue congestion. Thus NDN adjusts $cwnd$ mostly on Interest timeouts. Consequently, the $cwnd$ increases on every data received, ignoring potential loss in the network, while channel contention keeps getting higher, causing further packet loss. This effect exaggerates in a multi-consumer-producer scenario, leading to no queue congestion altogether. We consider this phenomenon as \textbf{\textit{cwnd} overestimation}, which significantly degrades NDN throughput, as we will see later in Sec~\ref{subsec:performance-cwl-dil}.

To mitigate this overestimation, we use a Congestion Window Limit (CWL) approach from \cite{chen2003settingCWL}. 
It uses the round-trip hop-count (RTHC) to impose a tighter upper bound on the $cwnd$ and minimizes channel contention. It also considers the possibility of an asymmetric path for data and \texttt{ACK}. In NDN, however, the Interest-Data path is symmetric, i.e., data strictly follows PIT entries downstream. Even with potential disjoint paths, the symmetric behavior between one Interest, one Data holds. Thus after a consumer receives a data packet and increases $cwnd$ using Eq.~\ref{eq:ndn-aimd-increase}, it can use the packet's \texttt{hop\_count} ($=\frac{RTHC}{2}$) from the data node to calculate the CWL ($cwl$) as follows,

\begin{verbatim}
if (hop_count <= 2) cwl = 2;
else if (hop_count <= 4) cwl = 1;
else if (hop_count <= 6) cwl = 2;
else if (hop_count <= 10) cwl = 3;
else if (hop_count <= 13) cwl = 4;
else if (hop_count <= 15) cwl = 5;
...
\end{verbatim}

Finally, we update the $cwnd$ with,
\begin{equation}
  cwnd = min(cwnd, cwl)
\label{eq:cwnd-cwl}
\end{equation}
Eq.~\ref{eq:cwnd-cwl} preserves congestion avoidance's linear growth and employs the tight upper bound to minimize channel contention. Simulation results in the next section also show significant throughput improvement using CWL in data-centric transport. 

\subsection{Effect of Interest Lifetime}
\label{subsec:dynamic-interest-lifetime}

Interest lifetime dictates how early or late an Interest will be evicted from the PIT. A small value means less aggregation, lower multicast opportunity, and vice-versa. Moreover, if a node receives a duplicate Interest with different \texttt{NONCE} before the existing PIT entry expires, it updates the PIT entry lifetime with the new Interest, increments the PIT count, and makes either one of two choices, (a) suppress, if it arrives within a suppression interval, or (b) forward. With RTO-based retransmission at the application level, case (b) will likely occur with a large and fixed Interest lifetime.

\begin{figure}[!t]
  \centering
  \subfloat[]{
    \includegraphics[width=0.30\columnwidth]{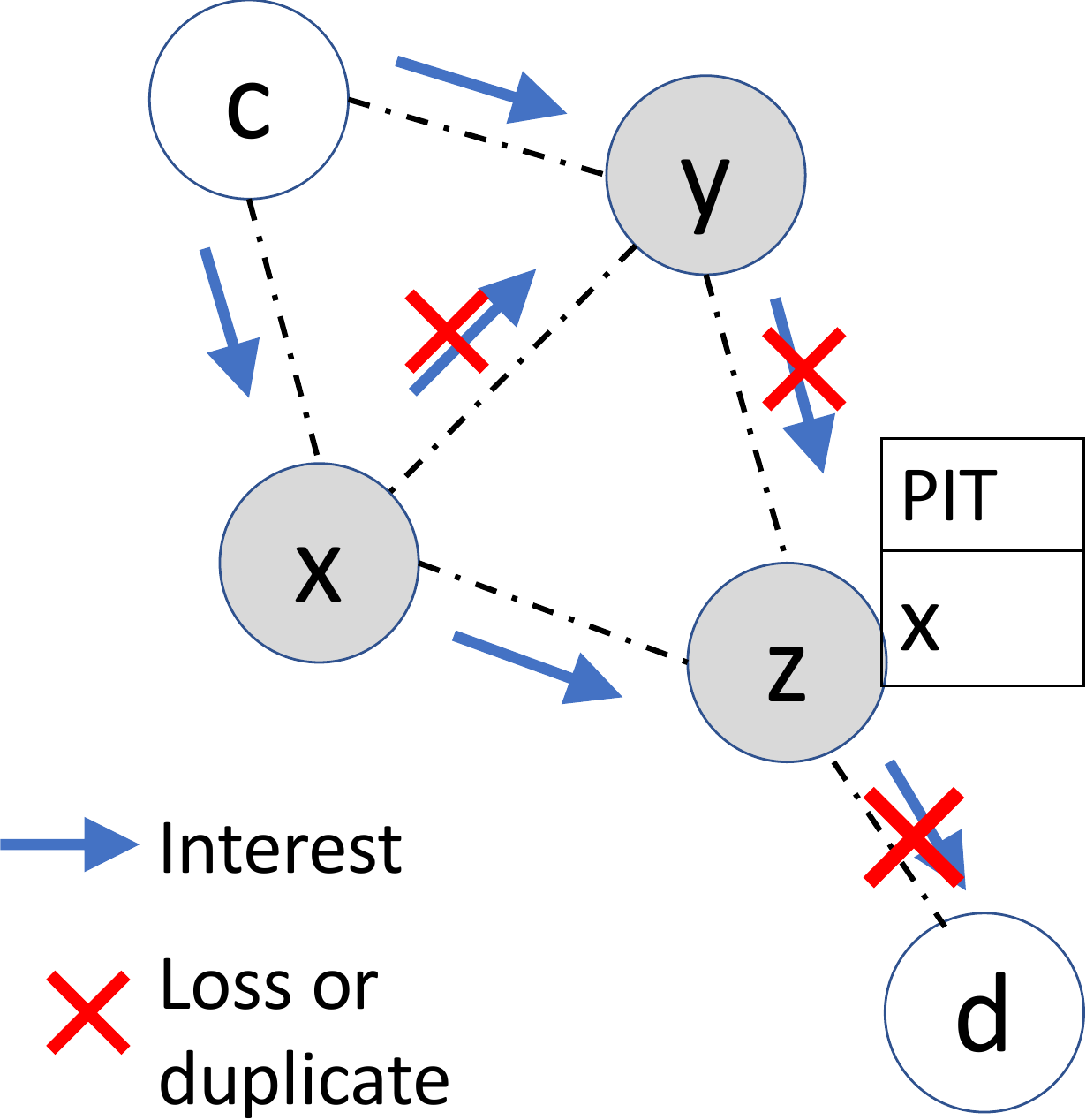}
    \label{fig:dil-first}
  }
  \hfill
  \subfloat[]{
    \includegraphics[width=0.30\columnwidth]{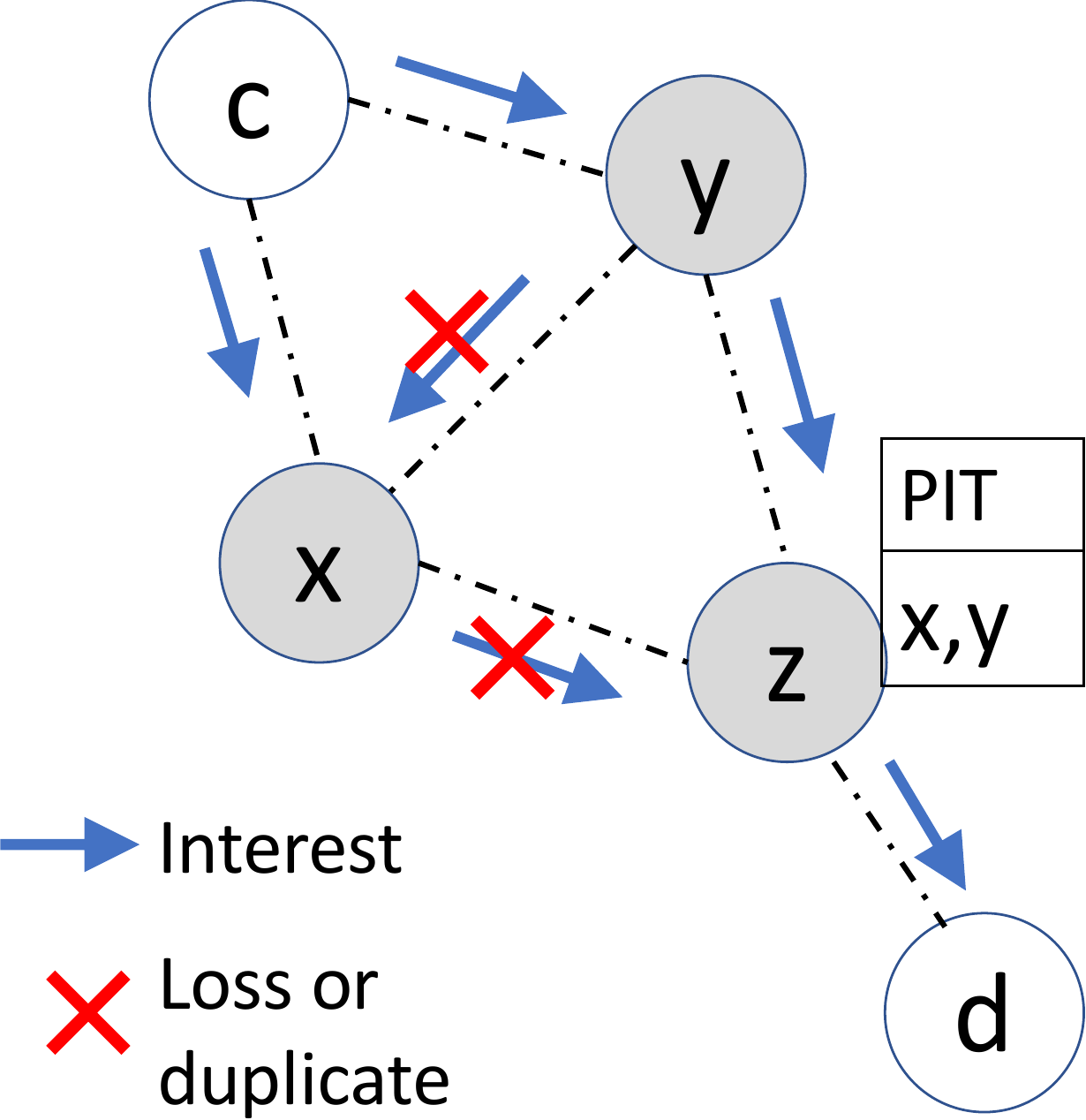}
    \label{fig:dil-second}
  }
  \hfill
  \subfloat[]{
    \includegraphics[width=0.30\columnwidth]{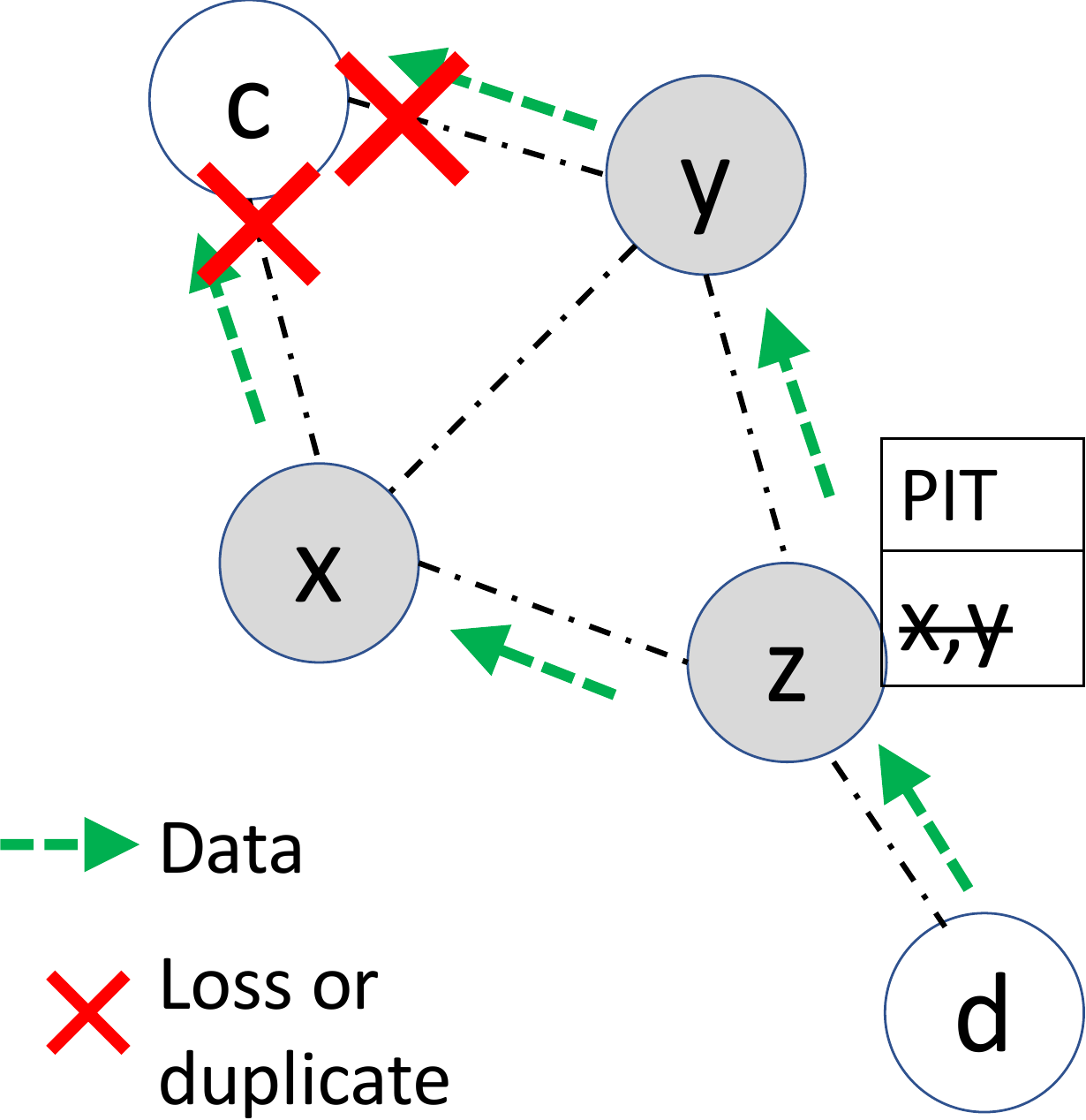}
    \label{fig:dil-third}
  }
  \caption{Example of data redundancy from multicast/broadcast on Interest aggregation upon application retransmission.}
  \label{fig:rtx-multicast}
\end{figure}

An example in Fig.~\ref{fig:rtx-multicast} shows $c$ as a consumer, $d$ as a data node, and the rest as forwarders. The dotted lines represent wireless reachability. Assume that $c$ broadcasts an Interest \texttt{/a/img.png} with 2s lifetime in Fig.~\ref{fig:dil-first}. Both $x$ and $y$ receive it, but $x$ broadcasts first from the layer-1 random timer during transmission, and $y$ detects a duplicate. Even though $x$ and $y$ both forward the Interest, the smaller packet size has a lower collision chance at $z$. Thus $z$ receives the Interest from $x$ and creates a PIT entry with $x$ as downstream (loop detected for $y$, dropped). However, the Interest fails to reach $d$ from collision/contention from another node nearby. At time=1s, the Interest times-out at $c$'s application and issues a retransmission (Fig.~\ref{fig:dil-second}). This time $z$ receives the Interest from $y$. Thus $z$ now has two downstreams, $x$ and $y$ for \texttt{/a/img.png}, forwards Interest to $d$ as the PIT considers case (b). Now, $d$ receives the Interest and sends the data to $z$ (Fig.~\ref{fig:dil-third}). Node $z$ will now either make two unicast transmissions as multicast or a single broadcast (according to the DAF strategy). Thus both $x$ and $y$ receive the data. As data packets are usually much larger than Interests, even with wireless layer-1 randomness, the chance of collision/contention at $c$ is highly likely by data transmission from both $x$ and $y$.

Thus, an extensive network can lead to a very high collision and contention. We use a Dynamic Interest Lifetime (\textbf{DIL}) technique to minimize the \textit{retransmitted Interest aggregation} effect. We use the most recent RTO (same as \cite{TCP-RTx}) as the subsequent Interest's lifetime and use a multiplier $\gamma > 1.0$ when checking for timeout events at the application as,
\begin{equation}
  timeout = RTO * \gamma.
\label{eq:dil-timeout}
\end{equation}

Using the calculated $timeout$ from Eq.~\ref{eq:dil-timeout} gives the existing PIT entry (or entries) for an Interest in the network a \textit{chance} to timeout before the application issues retransmission (RTx). It is because RTO over time varies, and at any given time, the calculated $timeout$ can be less than the assigned calculated lifetime of the most recent timed-out Interest. In our simulation, $\gamma=2.0$ offers the best results. Using an RTO-based lifetime and timeout checker lowers the RTx aggregation effect, proportional to $\gamma$'s value. Some redundant data from the RTx aggregation is still helpful when the network is mobile and sparse.

\subsection{Effect of In-network Caching}
\label{subsec:caching-effect}

\begin{figure}[!t]
  \centering
  \subfloat[]{
    \includegraphics[width=0.40\columnwidth]{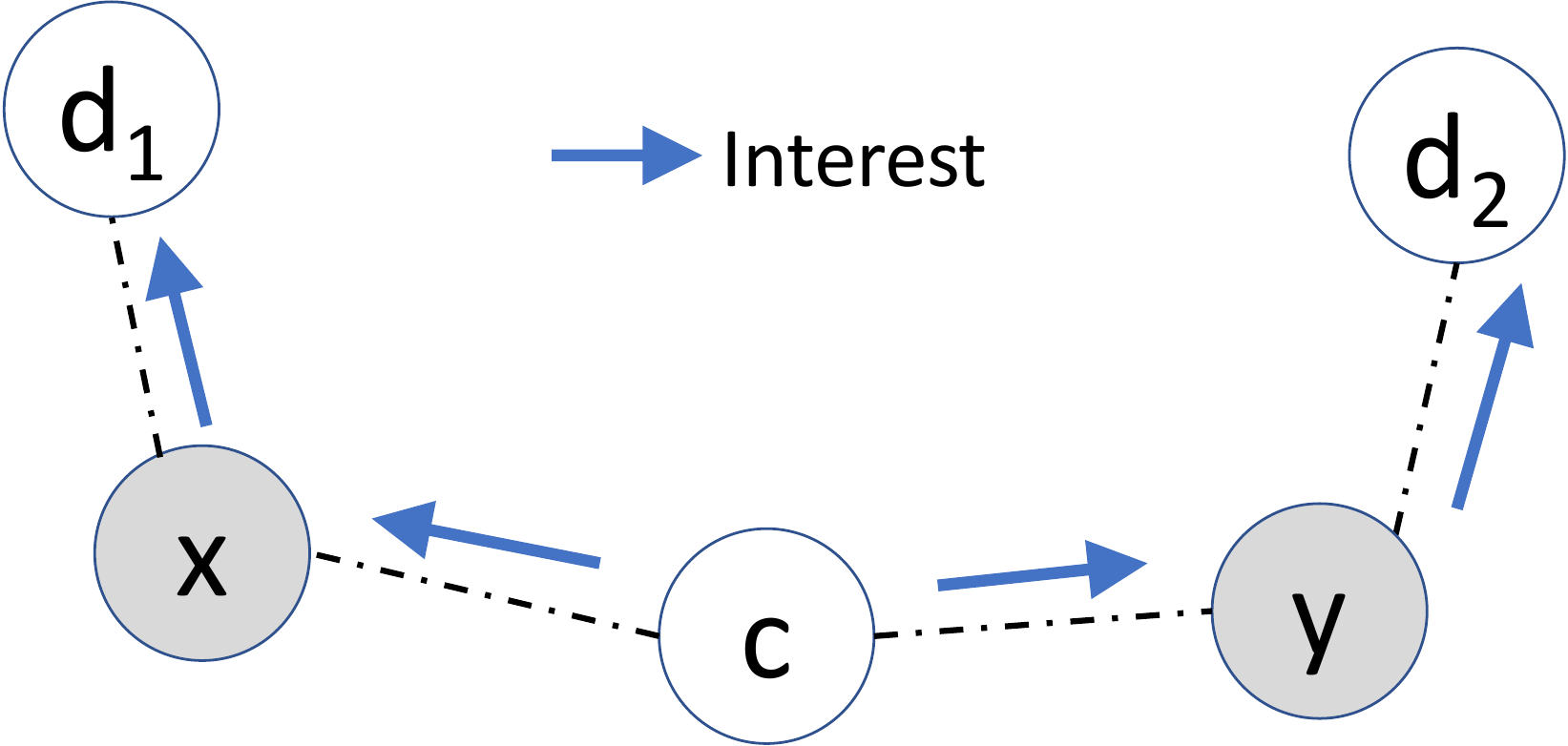}
    \label{fig:cache-first}
  }
  \medspace
  \subfloat[]{
    \includegraphics[width=0.40\columnwidth]{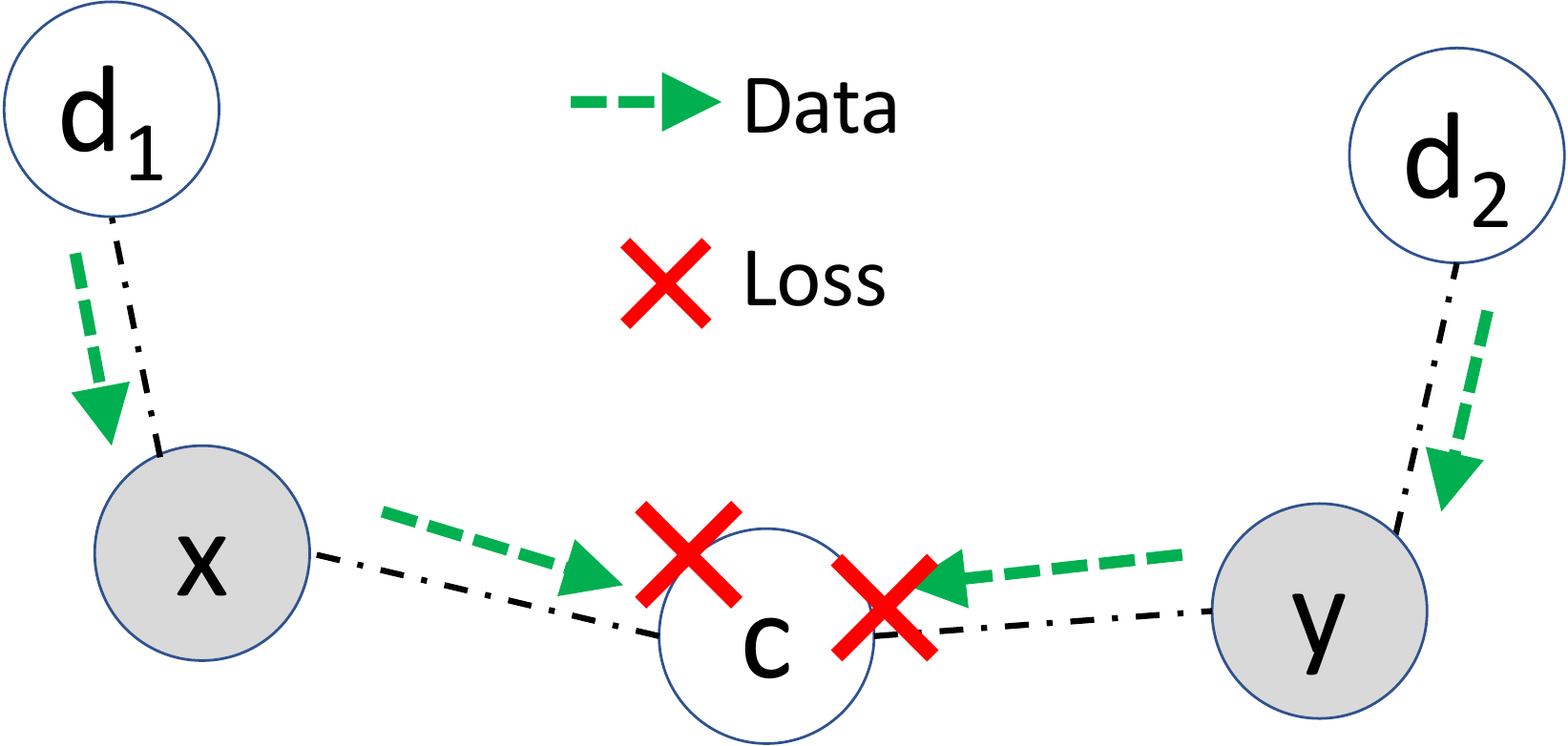}
    \label{fig:cache-second}
  }
  \caption{Example of data redundancy through caching.}
  \label{fig:cache-redundancy}
\end{figure}

The data-centric design enables in-network caching in NDN with content-store (CS), which can be highly effective under mobility and lossy networks in various ways. For example, if a data packet gets lost on its way to the consumer, a retransmission Interest may retrieve it from a potential closer cache node than reaching the actual producer. A cache hit can also occur if another consumer asked for the same-named data recently. However, caching can also lead to redundant data packets. In the example of Fig.~\ref{fig:cache-first}, we see two paths to data nodes $d_1$ and $d_2$ from consumer $c$ during the discovery phase. In case one of the paths breaks or suffers collision/contention, the other has the chance to retrieve data successfully. On the other hand, there is also a high probability of redundant data coming back to the consumer from both paths and causing a collision at $c$ (Fig.~\ref{fig:cache-second}). Thus caching can sometimes act as a double-edged sword depending on the network structure.

\section{Performance Analysis}
\label{sec:performance-analysis}
We now analyze the NDN AIMD performance in different setups of multiple consumer-producer communication with the discussed challenges and potential solutions in mind.

\subsection{Simulation Setup}
\label{sec:sim-setup}

We use ndnSIM~\cite{mastorakis2017evolution} to simulate the default NDN AIMD and the proposed CWL and DIL for reducing contention and redundant transmissions. We also use TCP-NewReno~\cite{tcpNewRenoRFC} for TCP/IP, with and without CWL. However, we do not consider added packet loss like in Sec.~\ref{sec:ndn-default-transport}, as multi-consumer-producer communication is good enough to induce packet loss from collision/contention. Each consumer or sender application runs for the entirety of the simulation duration, which is 100 seconds. RTS/CTS is disabled as a proper solution is not available for broadcast Interest and Data. Under mobility, all nodes move at the same speed in a single simulation. Each data packet payload is 512 bytes, and when DIL is disabled, the default Interest lifetime is 2 seconds. We also collect an average of 10 runs per simulation. Notations and values used in the plots and simulations are in Table~\ref{tab:notations}.

\begin{table}[!t]
  \caption{Notations and Values}
  \label{tab:notations}
  \centering
  \begin{tabular}{l|l|l}
    \hline
    \textbf{Notation} & \textbf{Meaning} & \textbf{Value}\\
    \hline
    cwl & CWL enabled & -\\
    \hline
    nocwl & CWL disabled (default AIMD) & -\\
    \hline
    cwl/dil & Both CWL and DIL enabled & -\\
    \hline
    $\gamma$ & Interest timeout checker multiplier & 2.0\\
    \hline
    CS & NDN content store capacity (packets) & 0 or 200 \\
    \hline
    1-1 & 1-to-1 communication & - \\
    \hline
    m-1 & many-to-1 communication & - \\
    \hline
    m-m & many-to-many communication & - \\
    \hline
  \end{tabular}
\end{table}

\subsection{Analyzing Channel Contention and Mobility}
\label{subsec:performance-cwl-dil}

\begin{figure}[!t]
  \centering
  \subfloat[Throughput.]{
    \includegraphics[width=0.47\columnwidth]{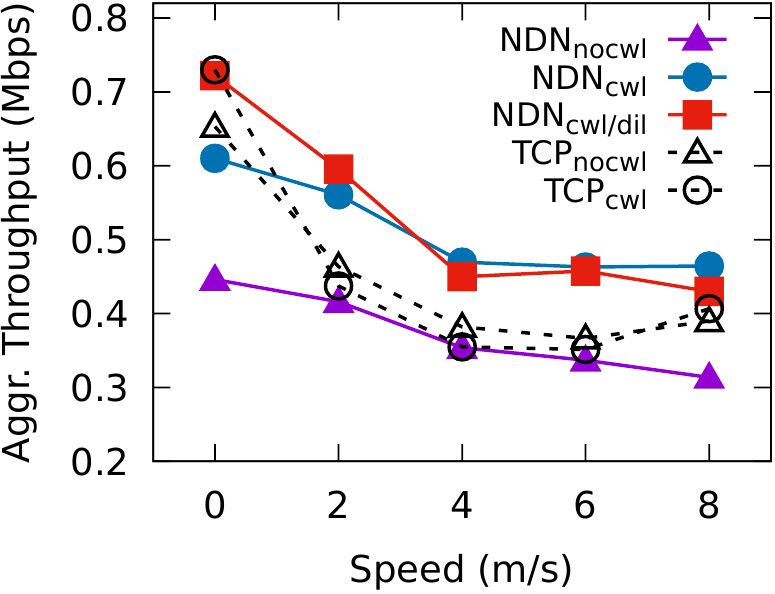}
    \label{fig:wl-throughput-cwl-dil}
  }
  \hfill
  \subfloat[Congestion window.]{
    \includegraphics[width=0.47\columnwidth]{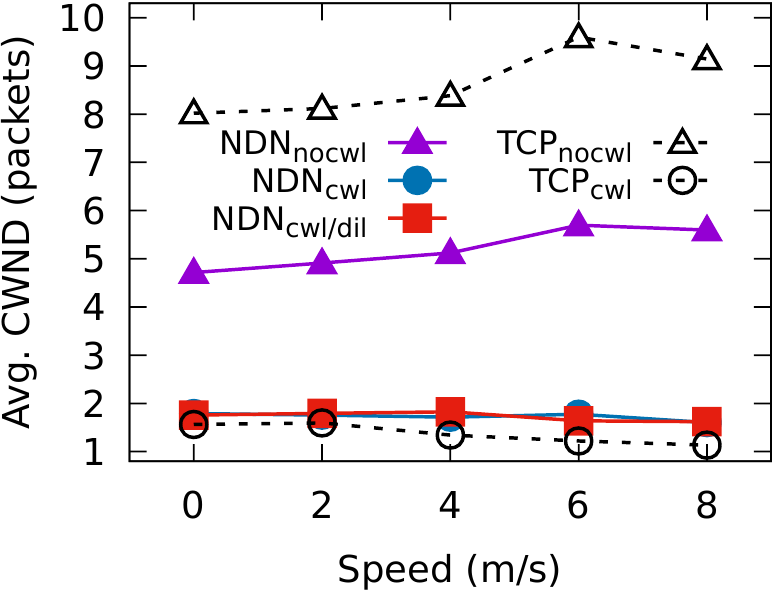}
    \label{fig:wl-cwnd-cwl-dil}
  }
  \par
  \subfloat[Round-trip time.]{
    \includegraphics[width=0.47\columnwidth]{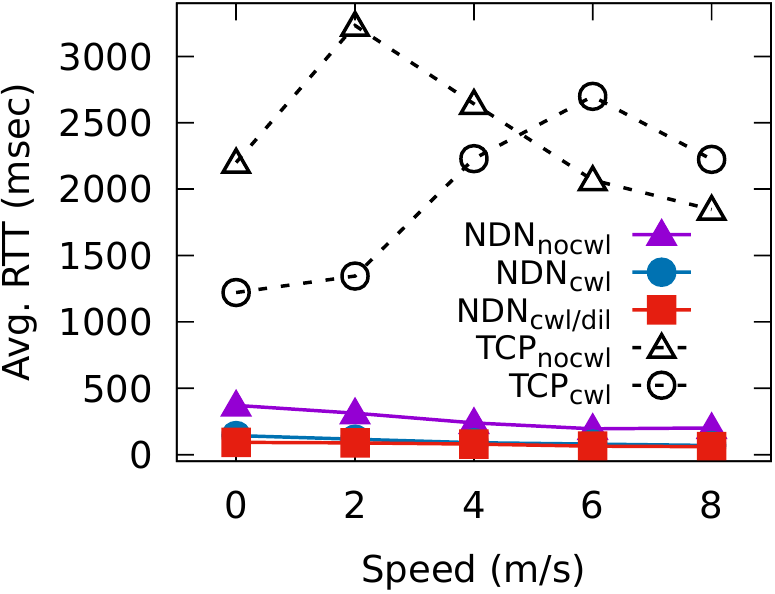}
    \label{fig:wl-rtt-cwl-dil}
  }
  \hfill
  \subfloat[Data broadcast events.]{
    \includegraphics[width=0.47\columnwidth]{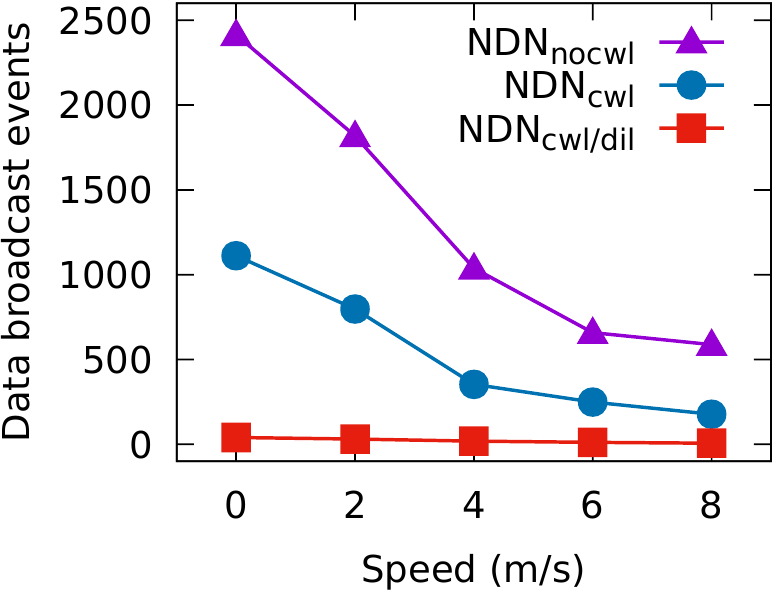}
    \label{fig:wl-dataBcast-cwl-dil}
  }
  \caption{Effect of channel contention, CWL and DIL on adaptive-rate application in 1-to-1 communication setup. CS=200 packets in NDN.}
  \label{fig:manet-transport-cwl-dil}
\end{figure}

We begin our analysis of the data-centric adaptive-rate application in wireless ad-hoc networks for multiple 1-to-1 consumer-producer communication, enabling and disabling mobility, in Fig.~\ref{fig:manet-transport-cwl-dil}. 
Here, each consumer-producer pair exchanges data under a unique prefix, i.e., each consumer retrieves data from only one specific producer and vice-versa. This setup shows the simplistic data-centric communication with out-of-order data retrieval. Moreover, multicast occurs only on the aggregation of retransmitted interest as different consumers ask for data under different name prefixes. TCP throughput inclusion is only for referential purposes. 

The aggregated throughput in Fig.~\ref{fig:wl-throughput-cwl-dil} shows the summation of application throughput in the ten consumers' applications. We can see that NDN has an average 17.25\% lower throughput than TCP when CWL is disabled, with the lowest 32.3\% when nodes are stationary. This result validates our claim that a default NDN AIMD suffers from $cwnd$ overestimation without CWL and DIL. With CWL alone, NDN's average throughput over different speeds exceeds 12.6\% compared to TCP with CWL. Note that TCP with CWL also has just about $<1\%$ throughput improvement over $nocwl$, on average, which is noticeable when nodes are stationary. It shows that CWL can help both TCP and NDN to lower channel contention in such scenarios.

However, a default Interest lifetime of 2 seconds still leads to unexpected Interest aggregation in the network on application retransmissions, leading to high data broadcast. Fig.~\ref{fig:wl-dataBcast-cwl-dil} verifies our claim showing $nocwl$ leads to the highest Interest aggregation, thus the highest data broadcast. The $cwl$ helps to reduce it to some extent, but not enough. Using both $cwl$ dynamic interest lifetime ($dil$) offers the best overall throughput in NDN (16.44\% more than TCP with $cwl$) as Data broadcast is the lowest, by allowing most in-network PIT entries to expire before retransmission. We further verify our claim that $cwl$ and $dil$ together reduce contention by looking into Fig.~\ref{fig:wl-cwnd-cwl-dil}, where high $cwnd$ represents high channel contention possibility and vice-versa.

Fig.~\ref{fig:wl-rtt-cwl-dil} further verifies that in NDN, lower RTT is a direct result of lower network contention and $cwl/dil$ offers the lowest values. It is also due to DAF's Interest-Data only communication avoiding any explicit routing exchange, unlike IP-AODV. The very high RTT over different speeds using TCP is mostly from AODV's route setup and maintenance. On average, NDN with $cwl$ and $dil$ offers an aggregated throughput of 0.53 Mbps while only $cwl$ and $nocwl$ yield 0.51 and 0.37 Mbps, respectively. Furthermore, $NDN_{cwl/dil}$ shows a maximum of 19.67\% more throughput than $NDN_{cwl}$, indicating that $dil$ works better in static networks.

\subsection{Analyzing Effects of Caching}
\label{subsec:performance-caching}

\begin{figure}[!t]
  \centering
  \subfloat[Throughput.]{
    \includegraphics[width=0.47\columnwidth]{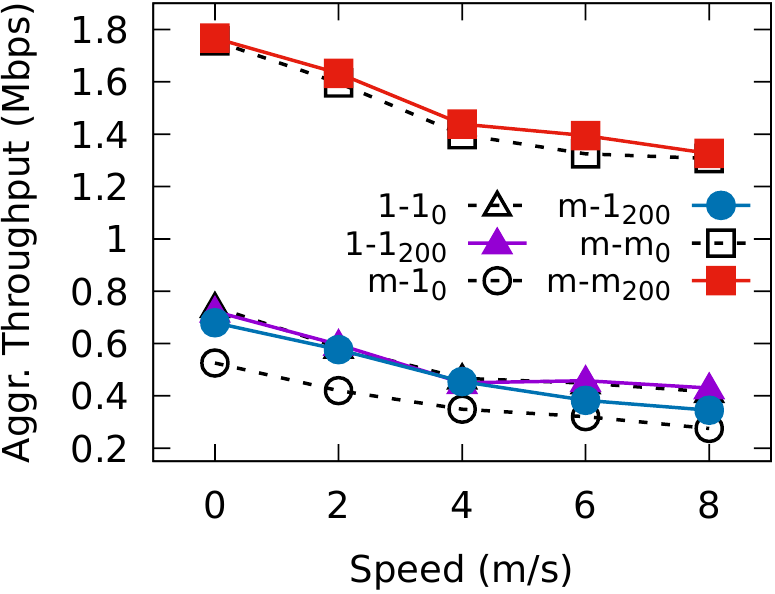}
    \label{fig:wl-cache-throughput}
  }
  \hfill
  \subfloat[Congestion window.]{
    \includegraphics[width=0.47\columnwidth]{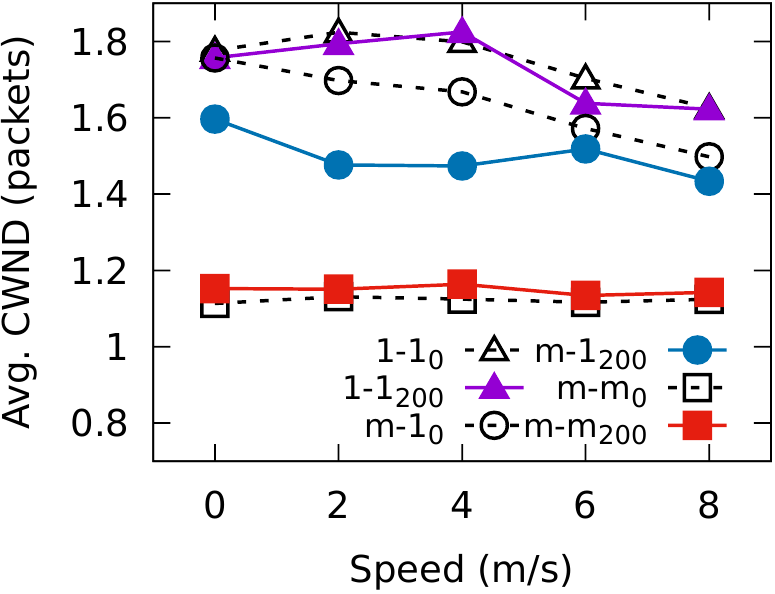}
    \label{fig:wl-cache-cwnd}
  }
  \par
  \subfloat[Round-trip time.]{
    \includegraphics[width=0.47\columnwidth]{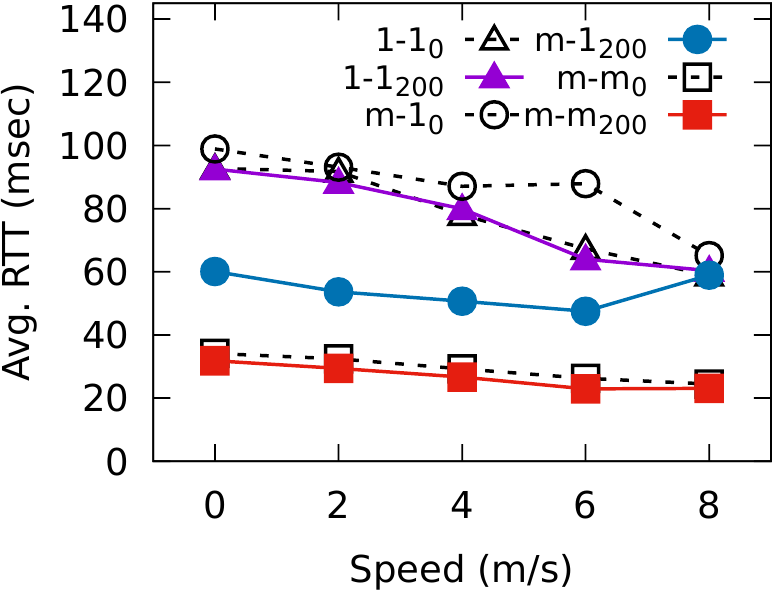}
    \label{fig:wl-cache-rtt}
  }
  \hfill
  \subfloat[Hop count.]{
    \includegraphics[width=0.47\columnwidth]{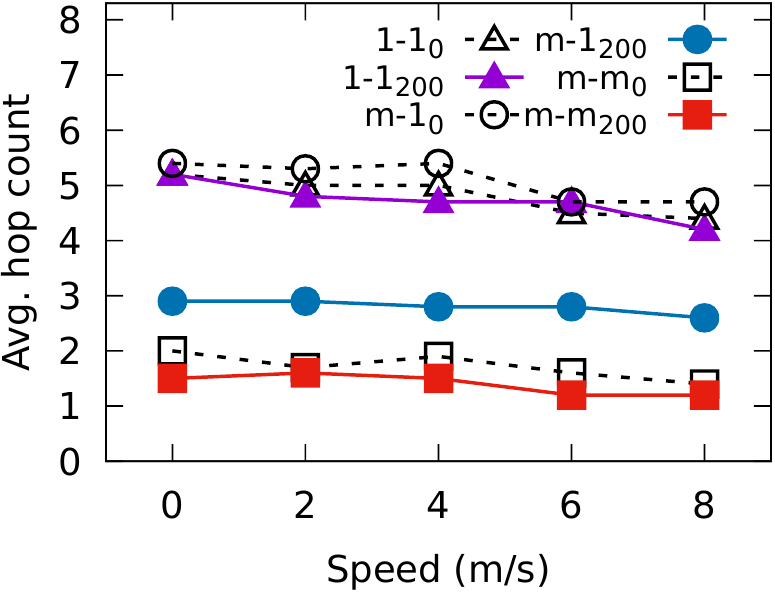}
    \label{fig:wl-cache-hops}
  }
  \caption{Effect of caching on adaptive-rate applications under different communication setups. Subscript shows per-node CS capacity in packets.}
  \label{fig:manet-transport-cache}
\end{figure}

Next, we analyze the effect of caching in NDN-based wireless ad-hoc networks with adaptive-rate applications and different communication setups in Fig.~\ref{fig:manet-transport-cache}, such as 1-to-1 (1-1), many-to-1 (m-1), and many-to-many (m-m). We do not show TCP results in the plots as IP does not support caching by default.

While 1-to-1 communication is the same as in Sec.~\ref{subsec:performance-cwl-dil}, in many-to-1 communication, there are two producers, serving data with two different prefixes, e.g., \texttt{A}, and \texttt{B}, respectively. Five consumers ask for the same set of data packets from one producer, and the other five consumers ask for the same set of data from the second one. In IP, however, one server sends five copies of the same data to five clients as multicast is not built-in to IP. This scenario expects the most NDN multicast, cache (when enabled) hits, and traffic reduction at (or near) an actual producer node. In many-to-many communication, five consumer-producer pairs exchange the same data set under the same prefix, e.g., \texttt{A}, while the other five pairs exchange data under another prefix, e.g., \texttt{B}. In IP, both 1-to-1 and many-to-many communications have the same behavior because of host-centric communication. However, in NDN, any eligible Data node (producer or cache) can serve a consumer following its location-agnostic design. Thus it shows the effect of retrieving data from the closest producer out of many by utilizing NDN multihoming and learning the shortest path towards data with DAF. All the communication schemes have $cwl$ and $dil$ enabled by default as they offer the best aggregated application throughput in our simulations. 

Fig.~\ref{fig:wl-cache-throughput} verifies our claim in Sec.~\ref{subsec:caching-effect} that the advantage of caching depends on the consumer-producer communication model. We see that only many-to-1 communication gains the most throughput with caching (28.83\% more than without). On the other hand, gains in the case of 1-to-1 and many-to-many are minimal. In fact, in a 1-to-1 scenario, caching results in about 0.4\% less throughput than without caching when nodes are stationary.

The caching effect is also visible in Figs.~\ref{fig:wl-cache-cwnd}, \ref{fig:wl-cache-rtt}, and \ref{fig:wl-cache-hops}, where many-to-1 and many-to-many schemes reduce the cwnd, RTT, and average hop count between consumer and data, respectively. Such behavior is expected as NDN's in-network caching helps bring data closer to the consumer under lossy and unstable conditions. Again, many-to-1 shows the highest RTT and hop count reduction as NDN multicast and caching reduce network load at and near the actual producer. Our dynamic interest lifetime does not hamper the performance for different consumers as $dil$ only reduces Interest aggregation on retransmissions from the same consumer using RTO and $\gamma$. However, in a 1-to-1 setup, $cwnd$ and RTT reduction is almost none because redundant data packets from caching end up increasing network transmission, and thus, collision and contention also increase. Fig.~\ref{fig:wl-cache-throughput} also shows that many-to-1 yields the lowest throughput without caching because of higher contention/collision possibility near a producer. 

One noticeable result is in many-to-many communication with caching, where the average aggregated throughput is 1.51 Mbps (Fig.~\ref{fig:wl-cache-throughput}), an average of 184.90\% more than the 1-to-1 setup with caching. The reason is NDN's location-agnostic design, where a consumer retrieves data from the closest producer or cache node. The lowest average hop count in Fig.~\ref{fig:wl-cache-hops} and the resulting low $cwnd$ and RTT in Fig.~\ref{fig:wl-cache-cwnd} and \ref{fig:wl-cache-rtt}, respectively, verify our claim.

Although we do not show the TCP/IP plots, on average, many-to-1 communication in NDN, with and without caching, shows 38.89\% and 7.66\% more throughput than TCP, respectively. In many-to-many communication with NDN, with and without caching, the throughput is on average 231.43\% and 223.69\% more than TCP. Such results reflect data-centric transport and forwarding advantages when it adequately handles $cwnd$ adaptation and Interest lifetime.

\section{Conclusion}
\label{sec:conclusion}

NDN's data-centric architecture can help achieve better throughput in adaptive-rate applications with and without loss in linear topologies than TCP/IP. However, we also find that in an extensive wireless ad-hoc network, NDN's out-of-order retrieval and improper Interest lifetime setting lead to mismanaged congestion window and redundant data. Consequently, high packet collision and channel contention degrade application throughput. We verify these effects by simulating synthetic data and an AIMD approach for NDN applications. We show that applying a congestion window limit and an RTO-based dynamic interest lifetime can significantly reduce the adverse effects. Together, they improve transport performance and show that one can achieve better throughput without changing the decentralized norm of NDN. We further show the effects of caching and that it is most beneficial when multiple consumers are asking for the same data. Finally, our analysis paves a path for future research on improvements and applications of data-centric architectures in real-world wireless ad-hoc networks. 

\bibliographystyle{IEEEtran}  
\bibliography{refs}

\end{document}